\documentclass[11pt]{article}
\usepackage{amsmath,amssymb,color,graphics,epsfig}

\usepackage{amsfonts,epsf,graphicx}

\makeatletter
\@addtoreset{equation}{section}
\makeatother

\usepackage{amsfonts}

\newcommand{\hoch}[1]{$\, ^{#1}$}

\newcommand{\be}{\begin{equation}}
\newcommand{\ee}{\end{equation}}
\newcommand{\bea}{\setlength\arraycolsep{2pt} \begin{eqnarray}}
\newcommand{\eea}{\end{eqnarray}}
\newcommand{\nn}{\nonumber}

\def\ft#1#2{{\textstyle{\frac{\scriptstyle #1}{\scriptstyle #2} } }}
\def\fft#1#2{{\frac{#1}{#2}}}

\def\0{{\sst{(0)}}}
\def\1{{\sst{(1)}}}
\def\2{{\sst{(2)}}}
\def\3{{\sst{(3)}}}
\def\4{{\sst{(4)}}}
\def\5{{\sst{(5)}}}
\def\6{{\sst{(6)}}}
\def\7{{\sst{(7)}}}
\def\8{{\sst{(8)}}}
\def\sst#1{{\scriptscriptstyle #1}}

\begin{document}

\begin{flushright}
\end{flushright}

\vspace{15pt}
\begin{center}
{\large {\bf Dynamic C-metrics in (Gauged) Supergravities}}

\vspace{10pt}
H. L\"u$^{1}$ and Justin F. V\'azquez-Poritz$^{2}$

\vspace{10pt}

\hoch{1}{\it Department of Physics, Beijing Normal University,
Beijing 100875, China}

\vspace{10pt}

\hoch{2}{\it Physics Department\\ New York City College of Technology, The City University of New York\\ 300 Jay Street, Brooklyn NY 11201, USA}

\vspace{30pt}

\underline{ABSTRACT}
\end{center}

We construct an exact time-dependent charged dilaton C-metric in four-dimensional ${\cal N}=4$ gauged supergravity. The scalar field drives the time evolution by transferring energy to the black holes, thereby causing their masses to increase and their acceleration to decrease. The values of the electric/magnetic and scalar charges determine three regions of potential time evolution. This solution holographically describes a strongly-coupled three-dimensional conformal field theory on the background of an evolving black hole. We also find new static charged dilaton C-metrics, which arise in four-dimensional Einstein-Maxwell-dilaton theories whose scalar potential can be expressed in terms of a superpotential.

\thispagestyle{empty}

\pagebreak



\newpage

\section{Introduction}

Gravitational solutions describing the formation and time evolution of black holes could help address a number of important issues in general relativity such as spacetime singularities, cosmic censorship and the production of gravitational waves (see recent reviews \cite{Joshi:2012mk,Gundlach:2007gc,Fryer:2011zz} and references therein). Moreover, according to the AdS/CFT correspondence, dynamical black holes in asymptotically anti-de Sitter spacetimes could provide a holographic description for certain non-equilibrium thermal systems of strongly-coupled field theories \cite{Aharony:1999ti}.

However, it is notoriously difficult to construct exact solutions that describe the time evolution of black holes. The most well-known solution that describes the formation of a black hole due to gravitational collapse is the Vaidya metric \cite{Vaidya:1951zz}. However, this involves only a generically specified matter energy-momentum tensor. Although they are rather rare, there are also exact dynamical solutions describing the formation of black holes in Einstein gravity coupled to fundamental matter fields such as scalars or vectors (for examples, see \cite{Roberts:1989sk,Gueven:1996zm}).

Recent progress has been made on this front. An exact time-dependent solution has been found that describes gravitational collapse to a static scalar-hairy black hole in four-dimensional Einstein gravity minimally coupled to a dilaton with a supergravity-inspired scalar potential \cite{Zhang:2014sta}. In addition, an exact collapse solution has been constructed in $D=4$, ${\cal N}=4$ gauged supergravity for which the gauge fields of the $U(1)\times U(1)$ subgroup of $SO(4)$ carry independent conserved charges \cite{Lu:2014eta}.

A key ingredient in these constructions is a non-conserved scalar charge which could be promoted to be a function of a null coordinate. It is therefore rather natural to consider other static solutions that involve scalar charges and might have analogous time-dependent generalizations. The candidates that we consider here are the dilatonic generalizations of the Einstein-Maxwell C-metric, which describe two oppositely charged dilaton black holes accelerating apart \cite{Dowker:1993bt}.

In section 2, we generalize the static dilaton C-metrics to be solutions of four-dimensional Einstein-Maxwell-dilaton theories with a scalar potential that can be expressed in terms of a superpotential. In section 3, we present a time-dependent C-metric in four-dimensional ${\cal N}=4$ gauged supergravity which describes dynamical black holes whose time evolution is driven by the dilaton. In section 4, we find that the boundary of the time-dependent solution has a black hole on it. This provides a holographic description of a strongly-coupled three-dimensional conformal field theory on the background of an evolving black hole. Section 5 provides a summary and discussion of our results.

\section{Static charged C-metrics}

Consider a family of four-dimensional theories given by
\be
e^{-1}{\cal L} = R-\ft12 (\partial\phi)^2-\ft14 e^{-a\phi} F_\2^2 - V(\phi)\,,
\ee
where $F_\2=dA_\1$ and the scalar potential can be expressed in terms of a superpotential as
\be
V = \left(\fft{dW}{d\phi}\right)^2-\fft34 W^2\,,\qquad
W = \fft{\sqrt{8}g}{a^2+1} \left( e^{\fft12 a\phi}+a^2 e^{-\fft{1}{2a}\phi}\right)\,.
\ee
These theories can be embedded in (gauged) supergravities for $a=\sqrt3,\ 1,\ 1/\sqrt3$ and $0$.  In particular, the $a=1$ case with $V=-2g^2 (\cosh\phi+2)$ arises from four-dimensional ${\cal N}=4$ gauged supergravity.

We find a solution given by
\bea\label{genastatic}
ds_4^2 &=& \fft{1}{(1+\alpha r x)^2} \Big[ f^{\fft{2a^2}{1+a^2}}\Big(- \Delta_r dt^2 + \fft{dr^2}{\Delta_r}\Big)
+ r^2h^{\fft{2a^2}{1+a^2}} \Big(\fft{dx^2}{\Delta_x} + \Delta_x d\varphi^2\Big)\Big]\,,\cr
e^{a\phi} &=& \left(\fft{f}{h}\right)^{\fft{2a^2}{1 + a^2}}\,,\qquad A_\1=4Q\, x\, d\varphi\,,\cr
\Delta_r &=& \widetilde\Delta_r h^{\fft{1-a^2}{1+a^2}}+ g^2 r^2 h^{\fft{2a^2}{1+a^2}}\,,\qquad
\Delta_x = \widetilde \Delta_x f^{\fft{1-a^2}{1+a^2}}\,,\cr
\widetilde\Delta_r &=& (1-\alpha^2 r^2) \left(1-\fft{2m}{r}\right)\,,\qquad
\widetilde \Delta_x = (1-x^2)(1+2\alpha m x)\,,\\
f &=& 1+ \alpha q x\,,\qquad h =  1 - \fft{q}{r}\,,\qquad m = \fft{2(a^2+1) Q^2}{q}\,.\nn
\eea
This describes magnetically charged black holes uniformly accelerating apart under the action of a cosmic string.
The vacuum AdS C-metric \cite{Plebanski:1976gy} can be recovered by taking $(Q,q)\rightarrow 0$ while keeping $m$ fixed. For the case of vanishing $g$, it was obtained in a different coordinate system in \cite{Dowker:1993bt}.
In our system, the compact coordinate $x\in [-\gamma,1]$, where $\gamma=\mbox{min}\left( 1,\ft{1}{2\alpha m}\right)$. The solution contains an additional hidden variable $C$, which describes the period of the coordinate $\varphi\in (-C\pi,C\pi)$.  One can choose $C$ appropriately in order to remove a conical singularity at either the north or south pole but not both. The global structure of the Ricci-flat C-metrics was analyzed in \cite{c-global}.

As with all C-metrics, the solution becomes more symmetric-looking if we let $r=-1/(\alpha y)$; however, it can be easier to extract the acceleration and black hole mass in the $(r,x)$ coordinates. For $g=0$, there is an event horizon located at $r=\mbox{min}\left( 2m,\alpha^{-1}\right)$ and an acceleration horizon at $r=\mbox{max}\left( 2m,\alpha^{-1}\right)$, giving rise to the Hawking and Unruh temperature, respectively. For non-vanishing $g$, both horizons are shifted due to contributions from the scalar potential. The magnetic charge is given by
$
Q_m=\ft{1}{16\pi} \int F_\2=C\,Q\,.
$
Unless it is essential, we do not always distinguish $Q_m$ and $Q$ in this paper. The black hole mass can be read off from the asymptotic falloff in $r$ with $\alpha=0$, giving
\be
M=C\Big(m + \ft{1-a^2}{2(1+a^2)} q\Big)\,.\label{mass}
\ee
For $a=1$, we can simply refer to $m$ as the mass of the solution.
The absence of a naked singularity for $g=0$ requires that $q\le 2\sqrt{a^2+1}\,Q$, which implies that $M\ge \ft{2}{\sqrt{a^2+1}}\,Q_{m}$, with the BPS (extremal) condition corresponding to the saturation of the inequality.  For non-vanishing $g$, the BPS and extremal conditions do not coincide.

   Finally, it is worth pointing out that we can take the Hodge dual of the 2-form field strength so that we have $e^{a\phi} \widetilde F_\2^2$ in the Lagrangian. The corresponding dual gauge field carries an electric charge, namely
\be
\tilde A_\1=\ft{4\tilde Q}{r}\, dt\quad\rightarrow\quad
\tilde Q_e=\ft{1}{16\pi} \int e^{a\phi} {\ast \widetilde F_\2} =C\, \tilde Q\,.\label{electric}
\ee

\section{Time-dependent C-metrics}

For $a=1$, we consider the time-dependent ansatz
\bea\label{time-dependent-solution}
ds_4^2 &=& \fft{1}{(1+\alpha(u) r x)^2} \Big[ f\left( -2\eta du dr - H(r,u,x) du^2\right)
+ r^2h \Big(\fft{dx^2}{\Delta_x} + \Delta_x d\varphi^2\Big)\Big]\,,\cr
e^{\phi} &=& \fft{f}{h}\,,\qquad A_\1=4Qx\ d\varphi\,,\\
\Delta_x &=& (1-x^2)\big( 1+2\alpha(u) m(u) x\big)\,,\qquad
f = 1+ \alpha(u) q(u) x\,,\qquad h = 1 - \fft{q(u)}{r}\,,\nn
\eea
where $\eta=\pm 1$ corresponds to the null coordinate $u$ being an advanced or retarded time. The electrically-charged ansatz associated with $e^\phi \tilde F_\2^2$ in ${\cal N}=4$ gauged supergravity is simply (\ref{electric}).  In the above ansatz, we let the effective mass $m$, acceleration $\alpha$ and the scalar charge $q$ depend on $u$ but the (conserved) $Q$ remains a constant.  The function $H$ is to be determined by the equations of motion.

We find three branches of solutions. The first has $\alpha(u)=0$, giving rise to
the single-charged version of exact black hole formation obtained in \cite{Lu:2014eta}.  For non-vanishing $\alpha$, we find that
\be
q(u)=q_0^2 \alpha(u)\,,\qquad m(u)=\fft{4Q^2}{q(u)}\,,
\ee
where $q_0$ is a constant, and the function $H$ is given by
\be
H = g^2r^2 h + (1 - \alpha^2r^2)\left( 1-\fft{2m}{r}\right)
- \fft{\eta\dot\alpha}{f^2}\Big(q_0^2 + 2 r^2 x + q_0^2 \alpha^2 r(3r-2q_0^2 \alpha) x^2\Big)\,.
\ee
Note that the function $\Delta_x$ remains independent of $u$. Constant $\alpha$ then gives rise to the previously-discussed static solution with $a=1$. On the other hand, a $u$-dependent $\alpha$ must satisfy
\be
\dot \alpha = \eta q_0^2 \alpha^{-2}\, (\alpha^4 - \alpha_1^4) (\alpha^2-\alpha_2^2)\,,\label{dotalpha}
\ee
where a dot denotes a derivative with respect to $u$ and
\be
\alpha_1 = \ft{2\sqrt2\,Q}{q_0^2}\,,\qquad \alpha_2=\ft{1}{q_0}\,.
\ee
The two fixed points $\alpha_1$ and $\alpha_2$ correspond to the masses
\be
m_1=\sqrt2\,Q \,,\qquad m_2=\ft{4Q^2}{q_0}\,.
\ee
Note that $m_1$ satisfies the BPS condition and depends only on the conserved charge $Q$. It follows from the inequality discussed under (\ref{mass}) that, for $a=1$, we have $8Q^2\ge q_0^2$ and hence
\be
\alpha_1\ge \alpha_2\,,\qquad m_1\le m_2\,.
\ee
This implies also that $x\in [-q_0^2/(8Q^2),1]$ for non-negative $\Delta_x$. Thus, for $\eta=+1$, if $\alpha = \alpha_1^-$ at $u= -\infty$ then we have $\dot\alpha<0$.  With the decrease of $\alpha$, $\dot\alpha$ increases to zero until $\alpha=\alpha_2^+$ at $u\rightarrow +\infty$.  The behavior and the relaxation times are
\bea
\alpha(u) \sim
\left\{
  \begin{array}{ll}
 \fft{2\sqrt2\,Q}{q_0^2} -c_1 e^{u/u_1}\quad & \hbox{for}\quad u\rightarrow -\infty \\
       \fft{1}{q_0} + c_2 e^{-u/u_2}\quad & \hbox{for}\quad u\rightarrow +\infty
  \end{array}
\right.
\eea
where
\be
u_1 = \ft{q_0^4}{8\sqrt{2}\, Q(8Q^2-q_0^2)}\,,\qquad u_2 = \ft{q_0^5}{2(64Q^4-q_0^4)}\,.
\ee
(Note that
$
u_1-u_2=\ft{q_0^4 (\sqrt{8}\, Q-q_0)^2}{8\sqrt{2}(64Q^4-q_0^4)}\ge 0\,.
$)
Therefore, we have the following picture: for $\eta=+1$, energy is transferred from the scalar field to the black holes, thereby increasing their mass from $m_1$ to $m_2$. Due to the increased mass, the acceleration of the black holes is lessened from $\alpha_1$ to $\alpha_2$.

For $Q\ne 0$, we can take $\alpha=0$ by also taking $q_0\rightarrow\infty$ such that $q$ is fixed, in which case we recover a static black hole solution. Thus, our time-dependent solutions must have $\alpha\ne 0$. For $\eta=+1$, if initially $\alpha<\alpha_2$ or $\alpha_2<\alpha<\alpha_1$, then $\alpha\rightarrow\alpha_2$ as $u\rightarrow\infty$. Thus, $\alpha_2$ is a stable fixed point for $\eta=+1$ and an unstable fixed point for $\eta=-1$. On the other hand, if $\alpha > \alpha_1$ initially for $\eta=+1$, then $\alpha\rightarrow\infty$ for a finite value of $u$. Thus, $\alpha_1$ is an unstable fixed point for $\eta=+1$, while it is stable for $\eta=-1$.

These two fixed points lead to three regions of evolution, as illustrated in Figure 1 for $q_0=1$ and $Q=1/\sqrt{2}$, which corresponds to $m_1=1$ and $m_2=2$. For each curve, we have chosen $\eta$ such that the effective mass increases rather than decreases with $u$. For the black curve, we have taken the initial mass to be $\epsilon=10^{-6}$ and we find that it evolves to the $m_1$ fixed point for $\eta=-1$. The red curve represents an evolution from an initial mass of $m_1+\epsilon$ to $m_2$ for $\eta=+1$. Lastly, the blue curve shows a mass that begins at $m_2+\epsilon$ and diverges for a finite value of $u$ for $\eta=-1$.
\begin{figure}[ht]
   \epsfxsize=3.0in \centerline{\epsffile{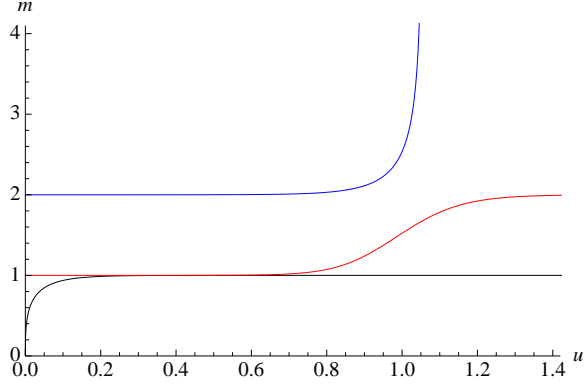}}
   \caption[FIG. \arabic{figure}.]{\footnotesize{Three types of time evolution of the effective mass $m(u)$.}}
\end{figure}

Figure 2 shows the evolution of the effective mass $m$ for $\eta=+1$, where we have taken the initial mass to be just above the $m_1=1$ fixed point for $Q=1/\sqrt{2}$ and we have considered the final fixed point to be $m_2=2$, $3$ and $4$ by adjusting the value of $q_0$. Interestingly, the larger the final mass, the less time it takes for the solution to evolve.
\begin{figure}[ht]
   \epsfxsize=3.0in \centerline{\epsffile{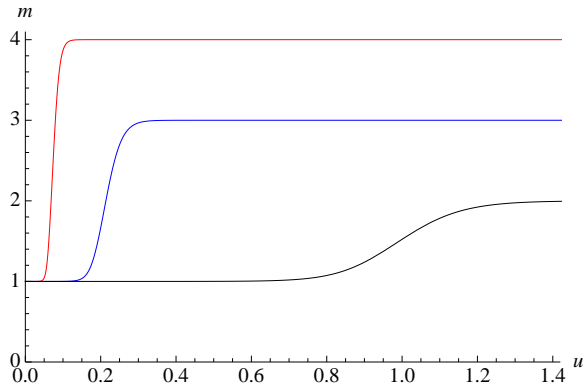}}
   \caption[FIG. \arabic{figure}.]{\footnotesize{The effective mass $m(u)$ for $m(0)=1$ and $m(\infty)=2$, $3$ and $4$.}}
\end{figure}

The solution (\ref{time-dependent-solution}) can be lifted to eleven dimensions using the reduction ansatz obtained in \cite{Cvetic:1999au} (see also \cite{Cvetic:1999xp}). The eleven-dimensional metric is given by
\bea
ds_{11}^2 &=& \Delta^{\fft23}\Bigg[ ds_4^2+\fft{4}{g^2}\Bigg( d\xi^2+\fft{1}{f\cos^2\xi+h\sin^2\xi}
\times \Big( h\cos^2\xi\ d\Omega_3^2+f\sin^2\xi\ d\tilde\Omega_3^2\Big)\Bigg)\Bigg]\,,\cr
d\Omega_3^2 &=& \big( d\psi+\cos\theta d\phi+\ft{1}{\sqrt{2}}gA_\1\big)^2+d\theta^2+\sin^2\theta d\phi^2\,,\cr
d\tilde\Omega_3^2 &=& (d\tilde\psi+\cos\tilde\theta d\tilde\phi)^2+d\tilde\theta^2+\sin^2\tilde\theta d\tilde\phi^2\,.
\eea
where the conformal factor is
$\Delta=(f\cos^2\xi+h\sin^2\xi)/\sqrt{fh}$.

\section{Black holes on the boundary}

Due to the conformal factor in the metric in (\ref{time-dependent-solution}), the boundary of the spacetime is at $1+\alpha rx=0$. Stripping off an overall conformal factor $f/(\alpha^2x^2(1+\alpha r x)^2)$, the boundary metric is given by
\be
ds_{\rm bdy}^2 =-2\eta\alpha\, du dx - F du^2 + \fft{dx^2}{\Delta_x} + \Delta_x d\varphi^2\,,
\ee
where
\be
\Delta_x=(1-x^2)f\,,\qquad
F=g^2 f -\alpha^2 \Delta_x - 2\eta x\,\dot\alpha\,,
\ee
and $f=1+x/x_0$ with $x_0=\ft{q_0^2}{8Q^2}<1$. This is a spatially compact universe with Killing horizons at the roots of $F$. According to the AdS/CFT correspondence, this describes a strongly-coupled three-dimensional conformal field theory on the background of an evolving black hole \cite{Aharony:1999ti}. This is a time-dependent generalization of the constructions in \cite{Hubeny:2009ru,Hubeny:2009kz}, where AdS C-metrics were used to describe field theories on a static black hole background.

Since the global structure of this evolving black hole is complicated, it is instructive to study the static limit at 
$\alpha=\alpha_{1,2}$. Then the metric takes the form
\be
ds^2=-F dt^2+ \fft{g^2 f}{F\Delta_x}\ dx^2 + \Delta_x d\varphi^2\,,\qquad dt=\eta du+\fft{\alpha}{F}\ dx\,.
\ee
Recall that there is a cosmic string in the bulk, which can be associated with conical defects on the boundary. We can choose the period $\Delta\varphi$ appropriately so that the metric is smooth at $x=1$. However, then there is a conical singularity at $x=-x_0$.  (For $x_0=\ft13$, both conical singularities at $x=-\ft13$ and $1$ can be avoided provided that $\Delta\varphi=\ft14\pi$.) The question is whether or not such a general conical singularity is naked.

For the $\alpha=\alpha_1$ static solution, there is a power-law curvature singularity at $x=x_0$, even though it also coincides with a Killing horizon. For sufficiently large $\alpha_1$, the singularity can be hidden inside a horizon at $x_1$ with $-x_0<x_1<1$. For the $\alpha=\alpha_2$ static solution, the solution describes a soliton with $x\in [-x_0,1]$ and with a conical singularity at $x=x_0$ except when $x_0=\fft13$.  For large enough $\alpha_2$, there is a horizon at $x_2$ with $x_2<x_1$. Thus, we see that as the bulk C-metric evolves from $\alpha=\alpha_1$ to $\alpha_2<\alpha_1$, the boundary metric evolves from a black hole to either another black hole or else a (smooth) soliton, depending on the parameters. The singularities, regardless of whether they are of the conical or power-law type, can be hidden inside a horizon for an appropriate parameter choice.

\section{Discussion}

We have presented two new types of C-metric solutions, which describe charged black holes accelerating apart due to the presence of a cosmic string. Firstly, we have found static charged C-metrics which arise in four-dimensional Einstein-Maxwell-dilaton theories whose scalar potential can be expressed in terms of a superpotential. Secondly, we have found a time-dependent C-metric solution in four-dimensional ${\cal N}=4$ gauged supergravity which describes black holes whose time evolution is driven by a scalar field. Namely, energy is transferred from the scalar field to the black holes, thereby causing their masses to increase and their acceleration to lessen. The magnetic (electric) and scalar charges specify three regions of time evolution.

The boundary metric of our time-dependent solution describes a three-dimensional black hole that evolves into either another black hole or else a soliton. For an appropriate choice of parameters, all conical and power-law type singularities are hidden inside a horizon. Via the AdS/CFT correspondence, this describes a strongly-coupled three-dimensional conformal field theory on the background of an evolving black hole. It would be useful to classify the dynamic black funnels and droplets contained within the bulk solutions, since this determines whether the black hole on the boundary is strongly or weakly coupled to the field theory \cite{Hubeny:2009ru,Hubeny:2009kz}. Moreover, it would be interesting to explore whether a black droplet can evolve into a black funnel, or vice versa.

AdS C-metrics have been used to construct localized black holes on a UV brane in the brane-world context \cite{Emparan:1999wa,Emparan:1999fd}, as well as plasma ball solutions on an IR brane \cite{Emparan:2009dj}. Our time-dependent solution might be used to describe a growing black hole on a UV brane or an expanding plasma ball on an IR brane. However, this could entail using a time-dependent version of the Israel junction conditions \cite{Israel:1966rt} to construct a dynamic two-brane.

It might be possible to generalize our time-dependent solution to be a solution of STU gauged supergravity with multiple electric and scalar charges. It would also be of substantial interest to find analogous dynamical solutions in higher dimensions, although at present it would seem that these constructions are unique to four dimensions.

\bigskip
\noindent{\bf Acknowledgements}:

\noindent The research of H.L. is supported in part by NSFC grants 11175269 and 11235003.

\end{document}